\documentstyle[11pt,twoside,jltp,amssymb]{article}

\title{Ferromagnetism and Superconductivity in Carbon-Based Systems}

\author{Y. Kopelevich\address{Instituto de F\'isica ``Gleb Wataghin", Universidade
Estadual de Campinas, Unicamp 13083-970, Campinas, S\~{a}o Paulo,
Brasil} and P. Esquinazi\address{Division of Superconductivity and
Magnetism, Institut f\"ur Experimentelle Physik II, Universit\"at
Leipzig, Linn{\'e}str. 5, D-04103 Leipzig, Germany}}

\runninghead{Y. Kopelevich and P. Esquinazi}{Ferromagnetism and
Superconductivity in Carbon-Based Systems}

\input psfig

\begin{document}

\begin{abstract}
In this article we shortly review previous and recently published
experimental results that provide evidence for intrinsic,
magnetic-impurity-free ferromagnetism  and for high-temperature
superconductivity in carbon-based materials. The available data
suggest that the origin of those phenomena is related to structural
disorder and the presence of light elements like
hydrogen, oxygen and/or sulfur.
\end{abstract}
\maketitle
\vspace{0.3in}

\section{INTRODUCTION}
In the last months of the year 1999 Frank Pobell, one of the editors
of the Journal for Low Temperature Physics,  received a manuscript
entitled ``Ferromagnetic- and Superconducting-Like Behavior of
Graphite" to be considered for publication. In this
manuscript\cite{1} we reported on the possible occurrence of
ferromagnetic as well as superconducting correlations in highly
oriented pyrolytic graphite (HOPG) samples at ambient conditions. It
is well known that graphite is the most diamagnetic material among
non-superconducting substances and it lacks d- and f- electrons,
which are generally assumed to be necessary for the occurrence of
ferromagnetism at relatively high temperatures ($T > 15$~K).
Materials that show superconducting properties at room temperature,
on the other hand, were (and still are) rather unknown. At such
circumstances, the skepticism expressed by the referees of the above
manuscript was understandable. The referees suggested that the
observed ferromagnetic behavior of graphite was governed by magnetic
impurities such as Fe, Fe$_3$O$_4$ or Fe$_2$O$_3$. In spite of the
referee skepticism, the editors of the journal decided to publish our
manuscript accompanied by the Editor's Note: ``There has been a
controversy between the opinions of the referees on the possible
influence of impurities on the observations. However the editors have
decided to publish this article because its content is of rather high
significance and may stimulate further work and discussion".
Certainly, we believe that the Editors decision was correct and
therefore we take this opportunity to report on recent as well as
rather unknown developments in this field, honoring Frank Pobell for
his work as Editor of the Journal of Low Temperature Physics.

\section{FERROMAGNETISM AND SUPERCONDUCTIVITY IN CARBON-BASED
MATERIALS: HISTORICAL NOTE}

\subsection{Ferromagnetism}
 Ferromagnetism at high temperatures in
carbon-based structures was reported in several tens of papers in the
last millennium\cite{2}. Unfortunately, a rigorous evaluation of
those studies is not an easy task because in most of them there is no
or a rather incomplete study of the impurity contents in the samples.
Nevertheless, there are a few publications that speak for an
intrinsic phenomenon. In 1987 Torrance et al.\cite{3} reported  that
the reaction of symmetrical triaminobenzene (C$_6$H$_9$N$_3$) with
iodine produces a black, insoluble polymer. This polymer showed, in
some of the runs, ferromagnetism up to 700~K, which is near its
decomposition temperature. Although some trace quantities of Fe were
found, neither its amount nor the observed irreversibility in the
magnetic moment as a function of temperature speak for magnetic
impurities as the possible source for the magnetism. Apparently, a
lack of reproducibility of the results reported by these authors
remained in the years to come.

It appears that in several of the studies published in the last
20~years the problem of reproducibility of magnetic carbon was a
general, major problem, which is not necessarily related to the
inclusion or not of impurities but also to a rather narrow window of
parameters necessary to produce the expected results. Even nowadays
and in spite of the improvement in the characterization methods, we
do not find yet a reliable method to produce magnetic order in carbon
with high reproducibility.

In 1991 Murata et al.\cite{4} measured the magnetization of
amorphous-like carbons (amorphous carbon has localized
$\pi$-electrons and its bonds are inconsistent with any other known
allotrope forms of carbon) prepared from tetraaza compounds (organic
monomers with different amounts of carbon, hydrogen and nitrogen) by
chemical vapor deposition method. The aza-carbon showed a
magnetization of 0.45~emu/g at room temperature and at 50~Oe applied
field. The saturation magnetization of the prepared films increased
as a function of the ratio between hydrogen and carbon (H/C) of the
starting material, up to values of the order of 10~emu/g (only a
factor 10 smaller than the magnetization at saturation in
magnetite)\cite{5}.

Murakami and Suematsu\cite{6} produced magnetic ordering in fullerene
crystals exposing them to light irradiation from a xenon lamp in the
presence of oxygen. They showed that the typical diamagnetism was
overwhelmed by a para- and ferromagnetic response after irradiation
of the sample under xenon light in oxygen for 2.5~hs. The decrease of
the magnetic moment (para- and ferro-magnetic contributions) after
annealing and its increase after leaving the sample in air for three
months speak against the contributions of Fe-impurities. A magnetic
moment of $0.1~\mu_B$ per C$_{60}$ molecule was estimated from the
separated magnetic part of the sample. The temperature dependence of
the saturation magnetization indicates an extraordinarily high Curie
temperature $T_c \sim$~800~K. Makarova et al.\cite{7} reported
recently similar effects for laser- and electron-beam-illuminated
C$_{60}$ films obtained either in air or oxygen-rich atmosphere.

\subsection{Superconductivity}
 Superconductivity in carbon-based
materials was first found in alkali-metal graphite intercalation
compound (GIC) C$_8$K with a superconducting transition temperature
$T_c = 0.15~$K\cite{8}. Till very recently (see below), the highest
$T_c = 5$~K was reached in the GIC C$_2$Na\cite{9}. Afterwards, the
research work on the superconductivity in carbon-based materials has
been mainly focused on fullerene-based compounds, triggered by the
observation of superconductivity in alkali-doped C$_{60}$
(buckminsterfullerenes) at 18~K in K$_3$C$_{60}$\cite{10} and at 33~K
in Cs$_x$Rb$_y$C$_{60}$\cite{11}. An indication for high-temperature
superconductivity in interhalogen-doped fullerenes has been reported
by Song et al.\cite{12}. The interhalogen-doped fullerenes were
prepared by using iodine monochloride (ICl) as a dopant. SQUID
magnetization measurements revealed signatures of superconductivity
with a superconducting transition temperature above 60~K. Apparently,
the results were not reproduced afterwards and therefore this work
remained rather unknown by the community. Antonowicz \cite{13} on the
other hand, reported on possible room-temperature superconductivity
in aluminum-carbon-aluminum (Al-C-Al) sandwiches. As in Ref.~12, the
study in Ref.~13 did not trigger a broad scientific interest. We note
that superconductivity in many carbon-based materials is a metastable
phenomenon and therefore a lengthy, systematic experimental work is
needed.

\section{FERROMAGNETIC ORDER IN CARBON STRUCTURES}
\subsection{Magnetic impurities}

Iron is usually the main magnetic impurity one finds in carbon-based
materials. Due to its relatively large para- and ferromagnetic
contributions, the measurement of its concentration is of main
importance. There are different methods to perform this measurement.
In Ref.~14 Particle Induced X-ray Emission (PIXE) with protons was
used to measure the magnetic impurities of different graphite
samples. The results showed that for samples with concentration of
Fe-impurities between $0.3~\mu$g/g to $19~\mu$g/g the magnetization
at 2~kOe - after subtraction of background diamagnetic contribution -
does not show any correlation with the Fe-concentration. The results
also indicate that the ferromagnetic-like hysteresis loops are weakly
temperature dependent between 5~K and 300~K. The assumption that such
a small amount of Fe distributed in the carbon matrix behaves
ferromagnetically is consistent neither with the observed temperature
dependence nor with the behavior observed in graphite samples with
much larger Fe concentrations\cite{14}. To test the accuracy of the
PIXE method used in Ref.~14, recent measurements have been performed
on similar HOPG samples using neutron analysis and X-ray
fluorescence. The three methods agree within experimental accuracy
and indicate that, for example, HOPG samples of ZYA grade from
Advanced Ceramics have a Fe concentration below $1~\mu$g/g $(\sim
0.25~$ppm). Because the magnetic moment due to the ferromagnetic part
of carbon-based samples is usually small, great care must be taken
with the measurement of the impurities in all steps of the sample
handling.

\subsection{Recent Reports} In the last five years there were
several publications that indicate the existence of magnetic order at
relatively high temperatures in carbon-based structures. One of the
most cited papers on magnetic carbon was published after Ref.~1 and
reports on magnetic studies of polymerized fullerene with a Curie
temperature of 500~K\cite{15}. Impurity measurements done on these
samples after the publication revealed, however, that they had a
considerable amount of Fe impurities \cite{16,17}, clearly exceeding
the one reported in the original publication\cite{15}. Nevertheless,
the measurement of magnetic-like domains in impurity free
regions\cite{18} left us with doubts whether the Fe impurities were
responsible for the measured magnetic moment. Recent experiments
performed on samples prepared with mixtures of Fe and fullerene under
high temperatures and high pressures\cite{19} indicate that the Fe
particles transform totally in cementite Fe$_3$C with a Curie
temperature of 500~K. These last results therefore leave no doubt on
the origin of the magnetic signals in the samples reported in
Ref.~15. It should be noted, however, that the room-temperature
ferromagnetism in pressure polymerized fullerenes was also reported
in Refs.~20,21, in hydrofullerite C$_{60}$H$_{24}$ in Ref.~22 and
photopolymerized fullerene powder and films in Ref.~7.

As mentioned in section 1.1., early literature on magnetism in carbon
structures suggests that disorder and probably hydrogen or other
light elements like oxygen play a role in the reported
ferromagnetism. In particular, the work in Refs.~4,5 suggests a
correlation between hydrogen concentration and magnetic order in
carbon. Proton irradiation provides the unique possibility to implant
hydrogen, to produce lattice defects in the carbon structure and to
have simultaneously a complete elemental analysis of the magnetic
impurities in the sample.

Protons in the MeV energy range have a penetration depth of several
tens of micrometers inside a carbon structure. The defect formation
process by high energy protons is a non-equilibrium athermal process
and it appears rather unlikely that ordered arrays of defects are
formed by migration of interstitial carbon atoms or vacancies,
perhaps with the exception of interstitials across the gallery. In
Ref.~23 a review of the published results is given. Although the
number of variable irradiation parameters is large (energy, fluence,
proton current, temperature and samples state and thermal coupling)
it is shown that with implanted protons of the order of $100~\mu$C or
larger, observable effects are registered with the SQUID. More
systematic studies including the use of different kinds of ions are
necessary to understand the role of different irradiation parameters
to trigger magnetic order in carbon structures.

Talapatra et al.\cite{24} reported that Nitrogen and Carbon
irradiation of nanosized diamond powder triggers magnetic order at
room temperature. Whatever the origin for the observed phenomenon -
in that study no impurity analysis was presented - the results
indicate that further studies of irradiation effects on the magnetism
of carbon-based as well as of other nominally non-magnetic materials
will appear in the future.

The studies done in Ref.~25 reveal that carbon films prepared by CVD
on stainless steel substrates reach magnetization values of the order
of $0.15~$emu/g at room temperature, comparable to those reported in
earlier studies\cite{4,5}. In that paper the amount of measured
impurities appears to be not enough to account for the absolute value
of the magnetic moments of the samples.

Aging as well as low-temperature annealing effects on the magnetic
properties might be treated as experimental evidence against the
metallic impurity magnetism in graphite and other carbon-based
structures. These aging effects were reported for
C$_{60}$H$_{24}$\cite{22}, proton irradiated carbon
structures\cite{23} and oxygen-driven ferromagnetism\cite{26}. The
oxygen effect on the magnetic properties of graphite has been
explored in Ref.~26. In these experiments, an activated graphite
powder was prepared by cutting and grinding an ultraclean graphite
rod at $T = 300$~K in oxygen atmosphere by means of a virgin diamond
saw blade. It is found that whereas the starting sample demonstrated
a diamagnetic (non-hysteretic) response, measurements performed on
the oxidized graphite powder revealed a pronounced ferromagnetic
signal. It has been also found that the ferromagnetism vanishes with
time after taking the sample out from the oxygen atmosphere,
suggesting that the ferromagnetism is triggered by the adsorbed
oxygen and not by a possible trace of magnetic impurities.

\section{SUPERCONDUCTIVITY IN CARBON SYSTEMS: RECENT STUDIES AND
OUTLOOK}

Local superconductivity with $T_c$ ranging from $\sim 7~$K to 65~K
has been observed in sulfur-graphite composites
(C-S)\cite{27,28,29,30,31}. The results indicate that the
superconductivity occurs in a small sample fraction, possibly within
filaments and/or at the sample surface. These observations suggest
that adsorbed foreign atoms on the graphite surface can trigger both
ferromagnetism and superconductivity. Besides, indirect evidence has
been obtained that there is an interaction between the ferromagnetic
and superconducting order parameters. In particular, it has been
observed that the superconductivity occurrence in C-S rotates the
ferromagnetic moment direction by $90^{\circ}$  confining it within
the graphite basal planes\cite{30}.

The occurrence of superconductivity in carbon nanotubes (CNT), which
are graphite sheets folded into a cylindrical shape, has been
demonstrated by M. Kociak et al. at 0.55~K\cite{32}. Other results
suggest that superconducting correlations in CNT can take place at
much higher temperatures, viz. at 20~K\cite{33} and possibly even at
room temperature\cite{34}. An unambiguous evidence for the
superconductivity with $T_c = 12~$K has been reported for multiwalled
CNT\cite{35}. In 2004, superconductivity with $T_c \sim 4~$K has been
discovered in heavily boron-doped diamond synthesized at high
pressure/high temperatures conditions\cite{36}. Superconducting
diamond films with higher $T_c (\sim 7~$K, onset) were produced by a
CVD method\cite{37}.

Some progress has been achieved also in the synthesis of
superconducting GIC.  The superconducting intercalated compounds
C$_6$Yb ($T_c = 6. 5$~K) and C$_6$Ca with $T_c = 11.5~$K have been
obtained in Refs.~38,39. Thus and in spite of previously obtained
superconductivity in graphitic systems with higher
$T_c$\cite{27,28,29,30,31}, further experiments with GIC look
promising.

In 1992 Agrait et al.\cite{40} published scanning tunneling
spectroscopy results obtained on graphite surfaces at $T = 4.2~$K
that revealed a gap ($\Delta$) in the electronic structure of the
order of 50 to 100~meV. The authors suggested that the gap originates
from single electron charging effects, i.e. Coulomb blockade. In
fact, multiple maxima, periodic in voltage, are seen in the DOS when
contamination between the tip and the surface exists. However, a
single maximum in the DOS was obtained. Interestingly, this kind of
curves are not only reproducible but occurs only in disordered
surface regions, which may indicate that either contamination
produces an artifact or that the structure of disordered graphite has
local properties similar to superconductors and/or strongly
correlated systems.
\begin{figure}[h]
\centerline{\psfig{file=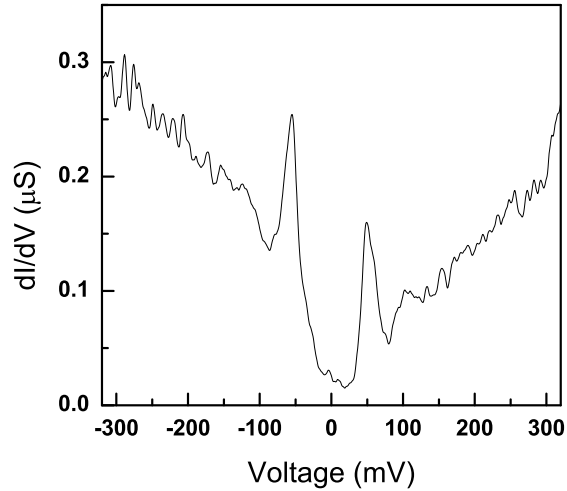,height=3.6in}}
\caption{``Anomalous" curve of differential conductance versus bias
voltage obtained at 4.2 K with a Au tip on a graphite surface. This
curve resembles that shown in Fig. 3(a) in Ref.~40.  The curve is
reproduced with permission of J. G. Rodrigo, Universidad Aut\'onoma
de Madrid. } \label{1}
\end{figure}

Figure 1 shows recent results of DOS measurements performed by one of
the authors of Ref.~40 on a pure graphite (HOPG) surface. Similar to
the results of Ref.~40 the gap signature in $dI/dV$ vs. $V$ is very
clear. It is tempting to conclude that well-defined peaks in the
tunneling conductance are related to coherence peaks expected within
the framework of the BCS theory for superconductivity. The value of
the bias voltage that corresponds to the peak of tunneling
conductance is a good measure for the superconducting gap
$\Delta_{sc}$. Then, the superconducting transition temperature
$T_{sc}$ can be estimated from the BCS equation
$2\Delta_{sc}/k_BT_{sc} \sim 3.5$. If the curve is due to a
superconducting surface region of graphite, it would indicate an
extremely large energy gap and consequently a very high critical
temperature. Using the data of Fig. 1 and the above equation, one
estimates $T_{sc} \sim 300 \ldots 400~$K. Alternative explanations
for the gap are also possible. For instance, the gap shown in Fig. 1
may also be related to the Kondo effect observed in other
carbon-based materials such as CNT \cite{N1,P1} and
C$^n_{60}$-molecule\cite{C}. Magnetic field dependent measurements as
well as measurements checking the state of the tip (by going back and
forth to a known metallic surface region like Au) are necessary to
understand the gap origin in graphite.

Finally, we would like to make a parallel between our studies\cite{1}
and earlier reports on the interplay between ferromagnetic- and
superconducting-like behavior of oxidized atatic polypropylene (OAPP)
and amorphous poly(dimethylsiloxane) (PDMS) polymers
samples\cite{41,42,43}. In both graphitic systems and the polymers,
the occurrence of either ferromagnetic- or superconducting-like
magnetization at room temperature depends on heat treatment,
oxidation, light illumination, and time (aging effect). Besides, in
the case of OAPP, a strong enough applied magnetic field could
transform the superconducting-like $M(H)$ to the ferromagnetic-like
$M(H)$ in an irreversible way, which has been attributed to the
field-induced breaking of diamagnetic (superconducting) loops,
resulting in a formation of ferromagnetic stripes\cite{41}.

Atatic polypropylene (C$_3$H$_6$) is a linear hydrocarbon where the
methyl groups are placed randomly on both sites of the chain. The
atatic chains are soft and rubbery, and can be easily oxidized.
Although not always measured, it is known that most of the studied
graphitic samples contain a large amount of hydrogen\cite{44}. So,
the formation of similar C-H-O structures responsible for the
magnetic (superconducting) behavior is possible in oxidized OAPP and
graphite.

Our previous work\cite{1} revealed that a low-vacuum heat treatment
of HOPG samples can either enhance the ferromagnetic response or
trigger superconducting-like $M(H)$ hysteresis loops at room
temperature, once again suggesting that adsorbed light elements play
a crucial role in the anomalous magnetic behavior of graphite. We may
speculate that the aging effect, i.e. the time dependence of the
sample magnetic response, is related to the oxygen
adsorption-desorption and/or its migration to different defect sites.
It is well known that graphite oxidation is triggered by the presence
of surface defects. This makes us believe that a combined effect of
structural disorder and adsorbed foreign atoms (molecules) such as S,
H, O (O$_2$) can be behind the anomalous magnetic behavior of
graphite and related carbon materials. Then, it is not unreasonable
to assume that aging effects, including the ferromagnetism
$\rightarrow$ superconductivity  transformation are related to
migration of foreign elements on the sample surface. The very small
$(\sim 0.01 \ldots 0.05$\%) Meissner fraction can also be understood
assuming the formation of superconducting patches (filamentary
loops\cite{43}) at the graphite surface. The accumulated experimental
evidence so far is certainly not enough to celebrate the discovery of
the superconductivity at very high temperature. Further experimental
work should concentrate on the production of the samples as well as
characterization and reproducibility of the effects reviewed in this
article.

\section{CORRIGENDUM TO THE ORIGINAL PAPER OF Ref.~1}

We take this opportunity to correct some erroneous data included in
the original publication\cite{1}. (a) In page 693 we have written
that the Fe concentration of our HOPG-2 sample was $90\pm 26~$ppm.
This value has been obtained by means of spectrographic analysis.
Measurements done on the same samples but with the PIXE analysis
revealed that the Fe concentration was much smaller, namely $2\pm
0.5~$ppm\cite{14}. The reason for this huge discrepancy is unknown.
Nevertheless, we stress that most of the HOPG samples we have
measured show an extremely low Fe concentration, much lower than the
values one finds in the literature. Because we have checked that our
analysis method is reliable, we speculate that upon the used method,
background contributions in the apparatus may give larger Fe
concentration when the sample mass or size is too small.

(b) The superconducting-like hysteresis loops obtained after
subtraction of the diamagnetic background and shown in Figs. 4, 5 and
7(a) in Ref.~1 are partially influenced by an artefact produced by
the SQUID current supply\cite{45}. This artefact is not due to the
superconducting solenoid but it depends in a non-simple way on the
maximum applied fields selected to measure the hysteresis.

\section*{ACKNOWLEDGMENTS}
The authors gratefully acknowledge discussions with Dr. R. H\"ohne,
Prof. S. Vieira, and Prof. S. Moehlecke. We thank Prof. J. G. Rodrigo
from the Universidad Aut\'onoma de Madrid for providing us with his
unpublished STM data taken on graphite, especially that shown in
Fig.1 of this manuscript and for stimulating discussions. This work
was supported by the DFG under Le 758/20-1, CNPq and FAPESP.

%\bibliographystyle{apsrev}
%\bibliography{D:/DATA/hopg/magnetic_carbon}   % name your BibTeX data base

\end{document}